# A Method for Estimating the Total Loss of Healthy Life Years: Applications and Comparisons in UK and Scotland

## Christos H Skiadas and Charilaos Skiadas

Technical University of Crete, Chania, Crete, Greece
E-mail: skiadas@cmsim.net
Hanover College, Indiana, USA
E-mail: skiadas@hanover.edu

**Abstract:** We propose a method of estimating the Total Loss of Healthy Life Years based on the first exit time theory for a stochastic process, the resulting Health State Function and the Deterioration Function estimated as the curvature of the health state function. We have done many applications in UK and Scotland and Sweden supporting our theory. Furthermore it was proven that both the WHO and EU estimates of the healthy life expectancy can result from our method. The WHO system takes into account the severe and moderate causes in estimating the loss of healthy life years; instead the EU system calculates the total loss of healthy life years. For both cases our methodology provides both estimators from only death and population data. The advantages of our method are straightforward. We do not need survey data to make the calculations. *The resulting estimates should be used to test and improve the existing survey based methodologies*. While the WHO and EU systems tend to approach each other differences continue to appear based on the methodology of the related surveys and the analysis of data. Two main schools are working to these directions one based on USA the Institute for Health Metrics and Evaluation (IHME) headed by Christopher J.L. Murray and contributors in all over the world and the European Health and Life Expectancy Information System (EHLEIS) with Jean-Marie Robine and a team from the EU member states.

## Introduction

The proposed method is based on the deterioration function presented in Figure 1. This is a function expressing the curvature of the health state function. The first part of this function, including the years from birth

---

until the expected healthy age (the age close to the maximum health state), covers the development stage during, which the human organism is growing, until the age of the maximum health, which corresponds to the minimum for the deterioration function. The second part of this function is related to the deterioration of the human organism following a growth pattern until a maximum deterioration in an age period ranging from 70 to 80 years nowadays and then declining at higher ages but, even so, continuing the deterioration of the human organism (see Skiadas 2011 Oct, 2011 Dec, 2012 Feb, Skiadas & Skiadas 2011 Jan, 2012 May).

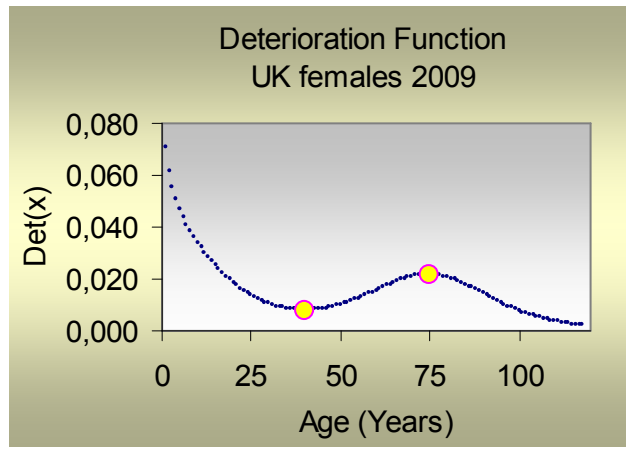

Fig. 1. The Deterioration Function

**The Method**

The task here is to estimate the influence of the deterioration process as a loss of the healthy years. This is achieved by the following formula:

$$TLHY = \int_0^x sDet(s)ds \approx \sum_0^x sDet(s)$$

Where the starting year denoted by $0 = x_{minDet}$ is at the minimum deterioration age and the final year denoted by $x = x_{H=0}$ is at the zero health state. The resulting estimator refers to the Total Loss of Healthy Years (TLHY). This estimator includes the loss of healthy years by Severe, Moderate and Light activity limitation. The years with severe activity limitation are estimated by a previously proposed estimator $LHLY_1$ (see Skiadas & Skiadas 2012 May), whereas the estimator $LHLY_3$ includes the severe and moderate activity limitation life years:

$$\text{Severe} + \text{Moderate} + \text{Light LHLY} = \text{Total LHLY}$$

$$LHLY_1 + (LHLY_3 - LHLY_1) + \text{Light LHLY} = \text{Total LHLY}$$

The estimates are done by using the latest version of the SKI-6-Parameters program in Excel which can be downloaded from the website: http://www.cmsim.net .

**Application in UK Females**

The results for females in UK 2009 are summarized in Table I. The age of the maximum deterioration is 75.1 years, the life expectancy at birth is 82.2 years and the healthy life expectancy at birth is 62.1 years. The total loss of healthy life is 20.1 years from which 7.7 years correspond to severe causes, 1.2 years to moderate and another 11.1 to light causes.

| \multicolumn{7}{c}{TABLE I} |
|---|---|---|---|---|---|---|
| \multicolumn{7}{c}{Healthy Years Estimates for UK, 2009 females} |
| Max Det Age | $LHLY_1$ Severe | $LHLY_3 - LHLY_1$ Moderate | Light LHLY | Total LHLY | HLEB | LEB |
| 75.1 | 7.7 | 1.2 | 11.1 | 20.1 | 62.1 | 82.2 |

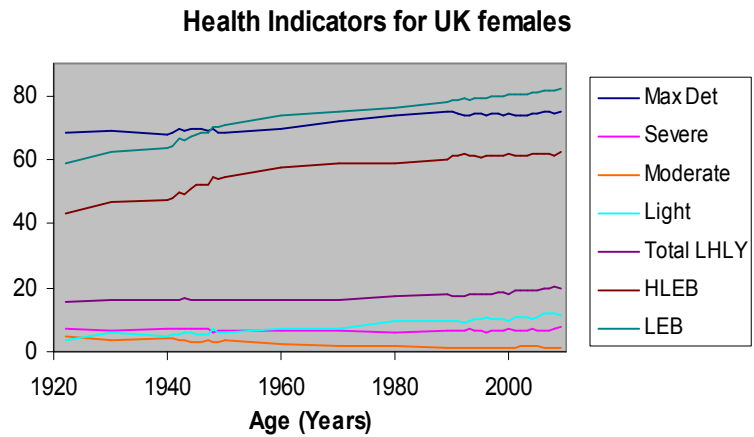

Fig. 2. Health Indicators for females in UK.

The total application for females in UK is summarized in Table II and the main findings are illustrated in Figure 2. The healthy life years at birth are growing during the last 9 decades but slower than the life expectancy at birth. This is because the total loss of healthy life years is higher by 4.6 years from 1922 to 2009 compared to the 23.5 years for the life expectancy at birth (LEB) for the same time period. In the same time the loss of healthy years due to severe causes was relatively stable, the moderate causes drop and the light causes grow considerably. The age of the maximum deterioration grew by 6.9 years from 68.2 in 1922 to 75.1 in 2009 verifying the relative stability of this indicator related to LEB.

**The EHEMU (EU) and HALE (WHO) estimates**

Another very important point is to close the gap between the healthy life expectancy estimates of the European Union based on the project EHEMU and the related estimates of the World Health Organization (WHO) under the code name HALE. As in both cases the estimates are based on different assumptions and differ considerably, the method used here provides information for both estimates. The HALE estimates of WHO give information for the expected healthy years by taking into consideration the severe and/or the moderate activity limitation, thus providing higher estimates for the expected healthy life years than the European Union Statistics. In the later case the estimates include part of what we refer to here as light activity limitation estimates as well. The EU (EHEMU, http://www.healthy-life-years.eu/) estimates are 64.79 years for males and 66.2 years for females for the healthy life expectancy at birth for UK (2007). Both estimates are within the range of our estimates including severe+moderate+part of light activity limitation as it is presented by the inequalities 59.4<65.0<69.2 and 61.6<66.2<73.4 for males and females respectively (the figures are from Table II and III). Our estimates for 2000 for males (67.3) and females (71.8) with severe & moderate activity limitation are within the uncertainty interval for the HALE estimates for 2000 that is 66.8<67.3<69.7 and 69.2<71.8<73.1 for males and females respectively. The HALE estimates for 2000 from WHO (annex4_en_HALE_2000) are 68.3 and 71.4 years for 2000 males and females, very close to our estimates.

| | | | TABLE II | | | | | | |
|---|---|---|---|---|---|---|---|---|---|
| Females UK | Max Det Age | $LHLY_1$ Severe | $LHLY_3 - LHLY_1$ Moderate | Light LHLY | Total LHLY | HLEB total | HLEB moderate &severe | HLEB severe | LEB |
| 1922 | 68,2 | 7,2 | 4,7 | 3,6 | 15,5 | 43,2 | 46,8 | 51,5 | 58,7 |
| 1930 | 68,7 | 6,6 | 3,6 | 5,8 | 16,0 | 46,6 | 52,4 | 56,0 | 62,6 |
| 1940 | 67,9 | 7,2 | 4,3 | 4,5 | 16,0 | 47,3 | 51,8 | 56,1 | 63,3 |
| 1941 | 68,5 | 7,1 | 4,0 | 5,2 | 16,2 | 47,7 | 52,9 | 56,9 | 63,9 |
| 1942 | 69,4 | 7,2 | 3,5 | 5,5 | 16,3 | 50,1 | 55,6 | 59,1 | 66,4 |
| 1943 | 69,1 | 7,1 | 3,8 | 5,9 | 16,8 | 49,4 | 55,4 | 59,1 | 66,3 |
| 1944 | 69,6 | 7,0 | 3,2 | 6,1 | 16,3 | 50,9 | 57,0 | 60,3 | 67,2 |
| 1945 | 69,3 | 7,5 | 3,2 | 5,3 | 16,0 | 52,1 | 57,4 | 60,6 | 68,1 |
| 1946 | 69,5 | 7,5 | 3,2 | 5,4 | 16,1 | 52,2 | 57,6 | 60,8 | 68,3 |
| 1947 | 68,8 | 7,2 | 3,6 | 5,7 | 16,4 | 51,9 | 57,6 | 61,1 | 68,3 |
| 1948 | 69,5 | 6,2 | 2,9 | 7,0 | 16,1 | 54,4 | 61,3 | 64,3 | 70,5 |
| 1949 | 68,3 | 6,7 | 3,3 | 6,0 | 16,0 | 54,1 | 60,2 | 63,4 | 70,2 |
| 1950 | 68,3 | 6,8 | 3,3 | 6,1 | 16,2 | 54,6 | 60,7 | 64,0 | 70,8 |
| 1960 | 69,7 | 6,8 | 2,4 | 7,0 | 16,2 | 57,5 | 64,5 | 66,9 | 73,7 |
| 1970 | 71,7 | 6,7 | 2,0 | 7,4 | 16,1 | 58,8 | 66,2 | 68,2 | 74,9 |
| 1980 | 73,7 | 6,2 | 1,7 | 9,6 | 17,6 | 58,9 | 68,6 | 70,3 | 76,5 |
| 1989 | 74,7 | 6,5 | 1,5 | 9,8 | 17,8 | 60,3 | 70,1 | 71,6 | 78,0 |
| 1990 | 75,1 | 6,6 | 1,5 | 9,4 | 17,5 | 60,9 | 70,3 | 71,8 | 78,4 |
| 1991 | 74,6 | 6,7 | 1,5 | 9,3 | 17,5 | 61,0 | 70,4 | 71,8 | 78,6 |
| 1992 | 74,0 | 6,8 | 1,5 | 9,1 | 17,4 | 61,6 | 70,6 | 72,1 | 78,9 |
| 1993 | 73,7 | 7,0 | 1,4 | 9,3 | 17,8 | 61,0 | 70,3 | 71,7 | 78,7 |
| 1994 | 74,2 | 6,6 | 1,5 | 10,0 | 18,1 | 61,1 | 71,2 | 72,6 | 79,3 |
| 1995 | 74,1 | 6,7 | 1,4 | 10,2 | 18,3 | 60,9 | 71,0 | 72,5 | 79,2 |
| 1996 | 73,9 | 6,2 | 1,4 | 10,5 | 18,2 | 61,2 | 71,7 | 73,2 | 79,4 |
| 1997 | 74,1 | 6,4 | 1,4 | 10,4 | 18,3 | 61,3 | 71,7 | 73,1 | 79,5 |
| 1998 | 74,1 | 6,5 | 1,5 | 10,3 | 18,3 | 61,4 | 71,7 | 73,1 | 79,7 |
| 1999 | 73,9 | 6,7 | 1,4 | 10,3 | 18,4 | 61,3 | 71,6 | 73,0 | 79,8 |
| 2000 | 74,2 | 7,0 | 1,4 | 9,9 | 18,2 | 61,9 | 71,8 | 73,2 | 80,2 |
| 2001 | 74,0 | 6,6 | 1,4 | 10,9 | 18,9 | 61,5 | 72,3 | 73,7 | 80,4 |
| 2002 | 73,9 | 6,3 | 1,8 | 10,9 | 19,1 | 61,4 | 72,3 | 74,1 | 80,5 |
| 2003 | 73,7 | 6,5 | 1,8 | 11,0 | 19,3 | 61,2 | 72,1 | 73,9 | 80,5 |
| 2004 | 74,4 | 7,0 | 1,7 | 10,4 | 19,1 | 61,9 | 72,3 | 74,0 | 81,0 |
| 2005 | 74,5 | 6,7 | 1,5 | 11,0 | 19,2 | 62,0 | 72,9 | 74,4 | 81,2 |
| 2006 | 74,7 | 6,6 | 1,4 | 11,8 | 19,8 | 61,6 | 73,5 | 74,9 | 81,4 |
| 2007 | 74,7 | 6,8 | 1,4 | 11,8 | 20,0 | 61,6 | 73,4 | 74,8 | 81,6 |
| 2008 | 74,6 | 7,2 | 1,3 | 11,8 | 20,3 | 61,3 | 73,1 | 74,4 | 81,6 |
| 2009 | 75,1 | 7,7 | 1,2 | 11,1 | 20,1 | 62,1 | 73,2 | 74,5 | 82,2 |

**Application in UK Males and Females**

The health indicators for males in UK are summarized in Table III. Figure 3 illustrates the life expectancy at birth (LEB) and the healthy life expectancy at birth for males and females in United Kingdom from 1922 to 2009. The immediate finding is that the life expectancy at birth is growing faster than the healthy life expectancy for both males and females. The mean healthy life expectancy for males for 20 years (1990-2009) is 59.8 years age whereas for females is 61.4 years of age. The trend is almost stable for males (0.0079) and slightly increasing for females. The trend for females is 0.042 corresponding to an increase of 0.8 years of healthy age at the 20 years period studied (see Figure 4). In the same period the increase in life expectancy at birth was 5.2 years for males and 3.8 years for females.

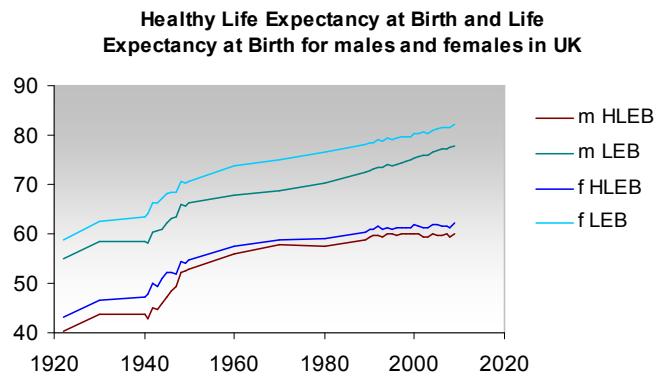

Fig. 3. Health indicators for males and females in UK

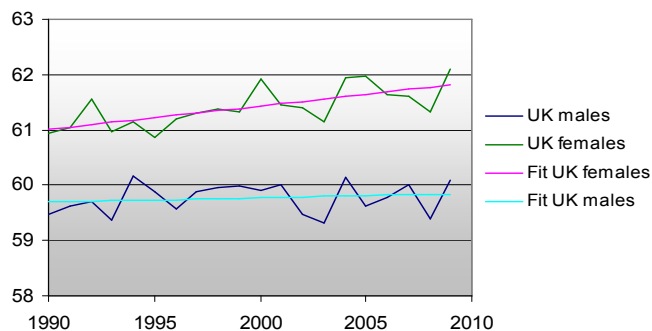

Fig. 4. The healthy life expectancy in UK for 20 years

It is important to note that the estimates based on surveys in different countries, cultures and economic systems vary considerably. Our method will provide a useful supporting tool to improve estimates and forecasts of the health indicators thus leading to a better future planning of the health system.

**TABLE III**

| Males UK | Max Det Age | $LHLY_1$ Severe | $LHLY_3$ - $LHLY_1$ Moderate | Light LHLY | Total LHLY | HLEB total | HLEB moderate &severe | HLEB severe | LEB |
|---|---|---|---|---|---|---|---|---|---|
| 1922 | 67,0 | 6,4 | 5,9 | 2,5 | 14,8 | 40,2 | 42,7 | 48,7 | 55,1 |
| 1930 | 67,3 | 6,2 | 4,5 | 4,0 | 14,7 | 43,7 | 47,7 | 52,2 | 58,4 |
| 1940 | 67,3 | 6,4 | 4,9 | 3,3 | 14,5 | 43,8 | 47,1 | 52,0 | 58,4 |
| 1941 | 68,3 | 6,7 | 5,0 | 3,6 | 15,3 | 42,9 | 46,5 | 51,5 | 58,2 |
| 1942 | 68,9 | 6,8 | 4,5 | 4,1 | 15,4 | 45,0 | 49,1 | 53,6 | 60,4 |
| 1943 | 69,0 | 7,0 | 4,7 | 4,1 | 15,7 | 44,7 | 48,8 | 53,5 | 60,5 |
| 1944 | 69,5 | 7,2 | 4,3 | 3,7 | 15,2 | 45,9 | 49,6 | 53,8 | 61,1 |
| 1945 | 69,6 | 6,7 | 4,4 | 3,7 | 14,9 | 47,2 | 51,0 | 55,4 | 62,1 |
| 1946 | 69,4 | 6,4 | 3,9 | 4,3 | 14,6 | 48,4 | 52,7 | 56,6 | 63,0 |
| 1947 | 67,7 | 6,5 | 3,8 | 3,7 | 14,0 | 49,4 | 53,1 | 56,9 | 63,4 |
| 1948 | 68,1 | 6,1 | 3,0 | 4,4 | 13,6 | 52,3 | 56,8 | 59,8 | 65,9 |
| 1949 | 66,8 | 6,2 | 3,1 | 4,0 | 13,3 | 52,5 | 56,5 | 59,5 | 65,7 |
| 1950 | 66,4 | 6,1 | 3,0 | 4,1 | 13,2 | 52,9 | 57,0 | 60,0 | 66,1 |
| 1960 | 45,5 | 5,7 | 2,8 | 3,5 | 12,0 | 55,9 | 59,3 | 62,2 | 67,9 |
| 1970 | 45,5 | 5,9 | 2,9 | 2,2 | 11,0 | 57,7 | 59,9 | 62,8 | 68,7 |
| 1980 | 64,1 | 6,0 | 2,7 | 4,2 | 12,9 | 57,6 | 61,8 | 64,4 | 70,5 |
| 1989 | 65,8 | 6,1 | 2,3 | 5,2 | 13,7 | 58,9 | 64,1 | 66,4 | 72,5 |
| 1990 | 65,9 | 6,2 | 2,3 | 4,8 | 13,3 | 59,5 | 64,3 | 66,6 | 72,8 |
| 1991 | 66,0 | 6,2 | 2,5 | 4,7 | 13,4 | 59,6 | 64,3 | 66,8 | 73,0 |
| 1992 | 66,2 | 6,0 | 2,3 | 5,4 | 13,8 | 59,7 | 65,2 | 67,5 | 73,5 |
| 1993 | 66,6 | 6,1 | 2,4 | 5,6 | 14,1 | 59,4 | 64,9 | 67,3 | 73,4 |
| 1994 | 67,9 | 6,1 | 2,2 | 5,5 | 13,8 | 60,2 | 65,7 | 67,8 | 73,9 |
| 1995 | 68,2 | 6,2 | 2,2 | 5,6 | 14,0 | 59,9 | 65,5 | 67,7 | 73,9 |
| 1996 | 69,1 | 5,6 | 2,6 | 6,4 | 14,6 | 59,6 | 66,0 | 68,6 | 74,2 |
| 1997 | 69,3 | 5,9 | 2,5 | 6,2 | 14,6 | 59,9 | 66,1 | 68,6 | 74,5 |
| 1998 | 69,7 | 6,1 | 2,5 | 6,1 | 14,7 | 60,0 | 66,1 | 68,6 | 74,7 |
| 1999 | 70,3 | 6,4 | 2,5 | 6,0 | 14,9 | 60,0 | 66,0 | 68,5 | 74,8 |
| 2000 | 70,9 | 5,9 | 2,2 | 7,4 | 15,4 | 59,9 | 67,3 | 69,5 | 75,3 |
| 2001 | 71,4 | 6,2 | 2,2 | 7,2 | 15,6 | 60,0 | 67,2 | 69,4 | 75,6 |
| 2002 | 71,8 | 6,0 | 2,0 | 8,3 | 16,3 | 59,5 | 67,7 | 69,7 | 75,8 |
| 2003 | 72,2 | 6,3 | 2,0 | 8,3 | 16,7 | 59,3 | 67,6 | 69,7 | 76,0 |
| 2004 | 72,7 | 6,8 | 1,9 | 7,7 | 16,4 | 60,1 | 67,9 | 69,8 | 76,6 |
| 2005 | 73,1 | 6,6 | 1,8 | 8,7 | 17,2 | 59,6 | 68,4 | 70,2 | 76,8 |
| 2006 | 73,6 | 6,6 | 1,7 | 8,9 | 17,3 | 59,8 | 68,7 | 70,4 | 77,0 |
| 2007 | 73,7 | 6,9 | 1,7 | 8,7 | 17,3 | 60,0 | 68,7 | 70,4 | 77,3 |
| 2008 | 73,9 | 6,7 | 1,6 | 9,8 | 18,1 | 59,4 | 69,2 | 70,8 | 77,5 |
| 2009 | 74,6 | 6,7 | 1,9 | 9,2 | 17,9 | 60,1 | 69,3 | 71,2 | 78,0 |

**The Health Indicators for Sweden**

The behavior of the health indicators during the pandemic influenza of 1918 was tested by using the data for Sweden from the Human Mortality Database (1910-1925, females). The results summarized in Table IV are illustrated in Figure 5. The resulting values for the age of the maximum deterioration remain relatively stable during this period as it is related to the main human mechanisms and not to the causes by influenza. Instead the healthy life years at birth (HLEB) and the life expectancy at birth (LEB) dropped considerably in 1918. The total loss of healthy life years was 2.4 years higher in 1918 than the previous year. More important changes came in the loss of healthy years from severe causes to 11.9 for 1918 from 8.7 years in 1917 and 8.2 than 5.1 for moderate causes respectively. Instead the light causes dropped from 4.7 years at 1917 to only 1 year the 1918.

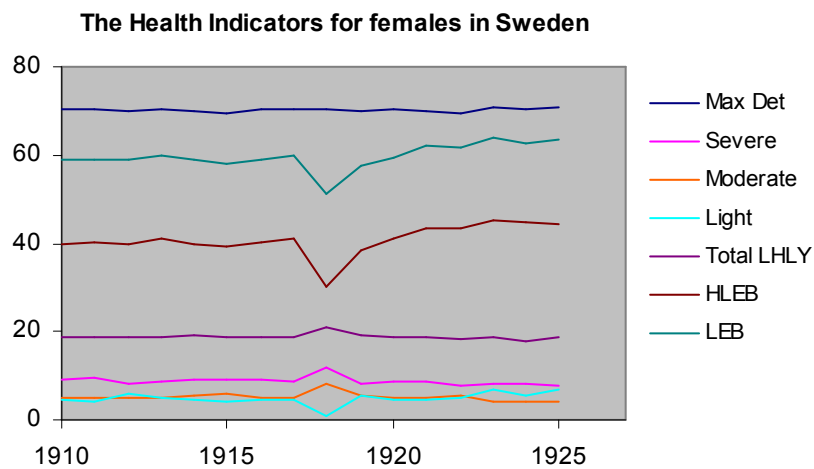

Fig. 5. The main health indicators in Sweden for the period 1910-1925.

TABLE IV

| Sweden | Max Det | Severe | Moderate | Light | Total LHLY | HLEB | LEB |
|---|---|---|---|---|---|---|---|
| 1910 | 70,3 | 9,2 | 5,0 | 4,6 | 18,8 | 40,0 | 58,8 |
| 1911 | 70,4 | 9,5 | 5,2 | 4,0 | 18,8 | 40,3 | 59,1 |
| 1912 | 69,8 | 8,1 | 5,0 | 5,8 | 18,9 | 39,9 | 58,8 |
| 1913 | 70,2 | 8,7 | 5,2 | 4,9 | 18,8 | 41,0 | 59,7 |
| 1914 | 70,1 | 9,1 | 5,5 | 4,6 | 19,2 | 39,9 | 59,1 |
| 1915 | 69,3 | 9,0 | 5,8 | 4,0 | 18,9 | 39,3 | 58,2 |
| 1916 | 70,4 | 9,1 | 5,1 | 4,4 | 18,6 | 40,4 | 59,0 |
| 1917 | 70,4 | 8,7 | 5,1 | 4,7 | 18,6 | 41,2 | 59,8 |
| 1918 | 70,5 | 11,9 | 8,2 | 1,0 | 21,0 | 30,2 | 51,2 |
| 1919 | 69,8 | 8,4 | 5,3 | 5,4 | 19,1 | 38,6 | 57,7 |
| 1920 | 70,2 | 8,5 | 5,2 | 4,8 | 18,5 | 40,9 | 59,5 |
| 1921 | 70,0 | 8,9 | 4,8 | 4,8 | 18,5 | 43,5 | 62,0 |
| 1922 | 69,7 | 8,0 | 5,4 | 4,9 | 18,2 | 43,5 | 61,7 |
| 1923 | 70,8 | 8,0 | 3,9 | 6,7 | 18,7 | 45,1 | 63,8 |
| 1924 | 70,2 | 8,2 | 4,1 | 5,7 | 18,0 | 44,8 | 62,8 |
| 1925 | 71,0 | 7,8 | 4,2 | 7,0 | 18,9 | 44,6 | 63,4 |

**Application to Scotland**

A quite good exploration of the healthy life expectancy in Scotland is done from 1980 to 2009. The applications are based on national and regional surveys and the data selected provided reliable estimates for the healthy life expectancy. Even more the estimates done include both the influence on life expectancy of the severe and/or moderate activity limitations mainly used in WHO and the total influence of all activity limitations accepted in European Union. As our method calculates both healthy life expectancy measures, we use the results from the application in Scotland to compare our direct system of estimates with the survey methods. Furthermore we expand our estimates starting from 1855 to 2009 based on the death and population data provided by the HMD for Scotland thus providing a useful estimate for the healthy life expectancy in Scotland during the last 155 years thus giving a powerful tool to policy makers to organize the health systems in our societies.

Two studies are of particular importance. The first study for Scotland is the 2004 paper for Healthy Life Expectancy in Scotland by Clark et al. 2006 and the second can be found from the official website: http://www.scotland.gov.uk/Topics/Statistics/Browse/Health/TrendLifeExpectancy , (12 July 2012 from the Scottish Government website). As in the previous case for UK we calculate the health estimates for Scotland for males and females. Table V summarizes the results for females. The last three columns of this table include the data from the two mentioned studies for Scotland. The first of these columns refers to the LLI (Limiting long-term term illness) a measure of serious and chronic ill-health which is closely related to our estimates referred as HLEB Total (The total Healthy life expectancy at birth). The second column (referred to as NEW method) is similar to the first while the third column (referred to as OLD method) is related to estimates without LLI. These estimates are related to our HLEB severe or HLEB moderate and severe.

| | | | | | | | | | | Scotland official estimates | | |
|---|---|---|---|---|---|---|---|---|---|---|---|---|
| Females Year | Max Det Age | $LHLY_1$ Severe | $LHLY_3 - LHLY_1$ Moderate | LHLY Light | LHLY Total | HLEB Total | HLEB moderate & severe | HLEB severe | LEB | HLE (LLI) | HLE New | HLE Old |
| 1855 | 68,2 | 6,8 | 5,3 | 0,8 | 12,9 | 31,3 | 32,0 | 37,4 | 44,2 | | | |
| 1880 | 68,7 | 7,0 | 4,9 | 1,7 | 13,7 | 31,7 | 33,4 | 38,3 | 45,4 | | | |
| 1900 | 70,3 | 6,5 | 5,5 | 1,3 | 13,3 | 34,8 | 36,1 | 41,6 | 48,1 | | | |
| 1910 | 69,9 | 6,6 | 5,0 | 2,8 | 14,5 | 38,3 | 41,2 | 46,2 | 52,8 | | | |
| 1920 | 70,0 | 6,8 | 4,4 | 4,8 | 16,0 | 39,0 | 43,8 | 48,2 | 55,1 | | | |
| 1930 | 68,9 | 6,7 | 4,5 | 4,9 | 16,1 | 43,1 | 48,0 | 52,5 | 59,2 | | | |
| 1940 | 67,2 | 6,8 | 4,5 | 4,0 | 15,3 | 45,8 | 49,8 | 54,3 | 61,1 | | | |
| 1950 | 68,8 | 6,3 | 3,6 | 6,2 | 16,1 | 51,9 | 58,1 | 61,7 | 68,1 | | | |
| 1960 | 69,1 | 6,4 | 2,7 | 6,7 | 15,8 | 56,1 | 62,8 | 65,4 | 71,8 | | | |
| 1970 | 71,5 | 6,7 | 2,3 | 6,9 | 15,9 | 57,5 | 64,4 | 66,6 | 73,3 | | | |
| 1980 | 75,0 | 6,5 | 1,2 | 8,8 | 16,6 | 58,6 | 67,4 | 68,6 | 75,1 | 61,0 | | 65,9 |
| 1981 | 75,1 | 5,8 | 1,6 | 8,9 | 16,3 | 59,1 | 67,9 | 69,6 | 75,4 | 60,6 | | 67,0 |
| 1982 | 74,5 | 5,7 | 1,9 | 9,1 | 16,7 | 58,5 | 67,7 | 69,5 | 75,3 | 60,5 | | 66,2 |
| 1983 | 74,8 | 6,0 | 1,7 | 8,8 | 16,5 | 59,2 | 68,0 | 69,7 | 75,7 | 61,5 | | 66,5 |
| 1984 | 75,5 | 6,6 | 1,3 | 8,3 | 16,3 | 59,6 | 67,9 | 69,2 | 75,9 | 61,1 | | 65,2 |
| 1985 | 74,8 | 6,0 | 1,5 | 8,4 | 15,9 | 59,9 | 68,3 | 69,8 | 75,8 | 61,4 | | 67,5 |
| 1986 | 74,4 | 6,1 | 1,6 | 8,5 | 16,1 | 60,1 | 68,6 | 70,2 | 76,3 | 60,8 | | 67,7 |
| 1987 | 74,9 | 6,2 | 1,4 | 8,0 | 15,5 | 61,0 | 68,9 | 70,3 | 76,5 | 59,0 | | 66,6 |
| 1988 | 74,8 | 5,8 | 1,4 | 9,1 | 16,3 | 60,4 | 69,5 | 70,9 | 76,7 | 59,8 | | 68,2 |
| 1989 | 73,9 | 6,4 | 1,4 | 8,2 | 16,0 | 60,1 | 68,3 | 69,7 | 76,1 | 62,3 | | 68,7 |
| 1990 | 75,4 | 5,9 | 1,5 | 8,9 | 16,3 | 60,6 | 69,5 | 71,0 | 76,9 | 61,1 | | 68,0 |
| 1991 | 73,9 | 6,5 | 1,3 | 8,6 | 16,4 | 60,7 | 69,3 | 70,6 | 77,1 | 61,9 | | 67,9 |
| 1992 | 73,6 | 6,2 | 1,2 | 8,3 | 15,8 | 61,5 | 69,8 | 71,0 | 77,3 | 61,3 | | 67,6 |
| 1993 | 72,8 | 6,7 | 1,4 | 7,5 | 15,7 | 61,3 | 68,8 | 70,3 | 77,0 | 59,7 | | 68,1 |
| 1994 | 76,1 | 6,1 | 1,3 | 9,3 | 16,6 | 61,0 | 70,3 | 71,6 | 77,7 | 60,5 | | 67,5 |
| 1995 | 75,0 | 6,7 | 1,2 | 8,0 | 15,9 | 61,8 | 69,8 | 71,0 | 77,7 | 60,1 | 59,3 | 67,8 |
| 1996 | 74,8 | 6,0 | 1,3 | 9,8 | 17,1 | 60,7 | 70,5 | 71,8 | 77,8 | 60,0 | | 69,1 |
| 1997 | 75,2 | 6,0 | 1,4 | 9,6 | 16,9 | 61,2 | 70,7 | 72,1 | 78,1 | | | |
| 1998 | 74,2 | 6,4 | 1,2 | 9,4 | 17,1 | 61,1 | 70,5 | 71,8 | 78,2 | 61,1 | 60,9 | 68,2 |
| 1999 | 73,7 | 6,3 | 1,3 | 9,7 | 17,3 | 61,0 | 70,6 | 71,9 | 78,2 | | | 67,7 |
| 2000 | 75,1 | 5,9 | 1,9 | 9,0 | 16,7 | 61,9 | 70,8 | 72,7 | 78,6 | 62,6 | | 68,7 |
| 2001 | 74,6 | 6,5 | 1,7 | 9,0 | 17,2 | 61,5 | 70,5 | 72,3 | 78,8 | | | 69,2 |
| 2002 | 74,2 | 6,8 | 1,7 | 9,0 | 17,6 | 61,2 | 70,3 | 72,0 | 78,8 | | | 69,3 |
| 2003 | 74,7 | 6,9 | 1,7 | 9,1 | 17,8 | 61,1 | 70,2 | 71,9 | 78,8 | | 60,5 | |
| 2004 | 75,1 | 6,4 | 1,4 | 10,4 | 18,1 | 61,2 | 71,6 | 73,0 | 79,4 | | | |
| 2005 | 75,4 | 6,4 | 1,5 | 10,4 | 18,2 | 61,2 | 71,6 | 73,0 | 79,4 | | | 69,4 |
| 2006 | 75,4 | 6,7 | 1,2 | 10,1 | 18,1 | 61,6 | 71,8 | 73,0 | 79,7 | | | 69,9 |
| 2007 | 75,0 | 7,4 | 1,1 | 10,3 | 18,7 | 61,0 | 71,3 | 72,4 | 79,7 | | | 70,2 |
| 2008 | 75,4 | 6,9 | 1,3 | 10,0 | 18,2 | 61,7 | 71,6 | 72,9 | 79,9 | | 62,4 | 70,8 |
| 2009 | 75,3 | 7,2 | 1,3 | 10,8 | 19,3 | 61,1 | 71,9 | 73,1 | 80,4 | | 62,2 | |

TABLE V (Application in Scotland for females)

Comparisons are illustrated in Figure 6. Our estimates for HLEB total, based on the total LHLY estimates are presented by the blue line in the graph. Also two confidence intervals at +-2 years are indicated by the dashed lines. The estimates based on surveys as the official estimates for Scotland should decline from the real situation by 2 to 3 years and we expect that will be included into this interval. The confidence interval for our estimates is negligible. The orange and the brown curves representing the HLE (LLI) and HLE (New) official estimates are included into the confidence intervals thus verifying a quite good approach to our direct estimates. The HLE (Old) official estimates (magenta line) are close to our estimates for HLEB with severe and moderate causes (green line). The main part of these data points are inside the +-2 confidence intervals. However, almost all the estimates for HLE (Old) are lower than the corresponding HLEB estimates.

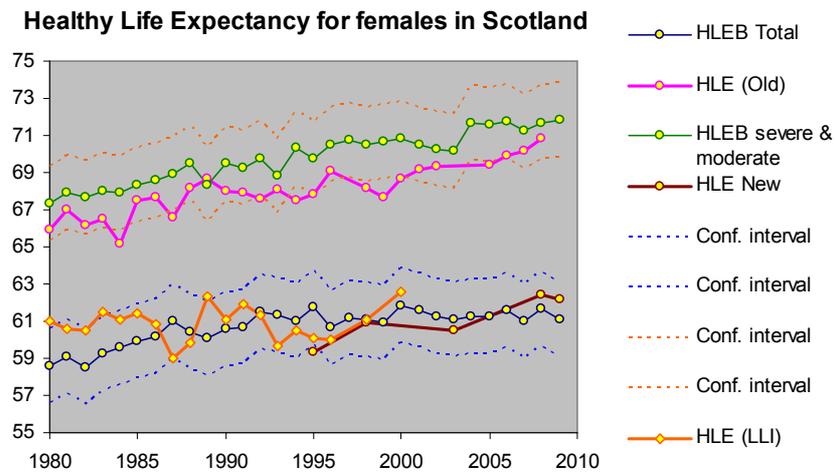

Fig. 6. Comparisons for females

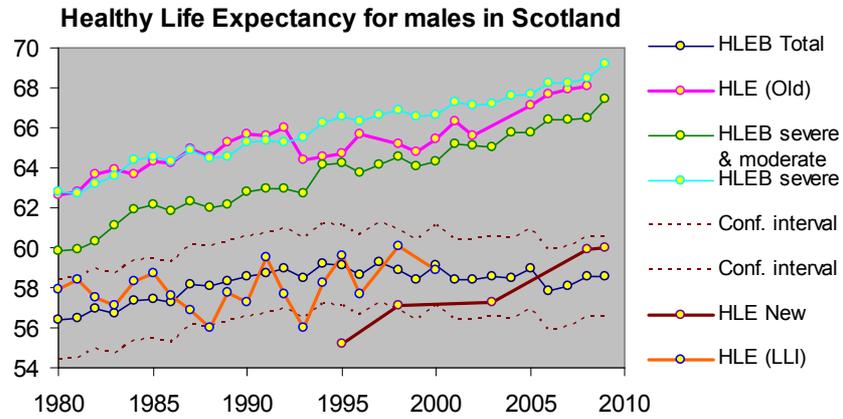

Fig. 7. Comparisons for males

As for females, comparisons for males in Scotland are illustrated in Figure 7. Table VI summarises our estimates and also includes the official estimates for males in Scotland. The estimates for HLEB total, based on the total LHLY estimates are presented by the blue line in the graph. Also two confidence intervals at +-2 years are indicated by the dashed lines. The orange and the brown curves representing the HLE (LLI) and HLE (New) official estimates are included into the confidence intervals thus verifying a quite good approach to our direct estimates. The HLE (Old) official estimates are close to our estimates for HLEB with severe causes (light bleu curve) and HLEB with severe and moderate causes (green curve).

*A Method for Estimating the Total Loss of Healthy Life Years* 13

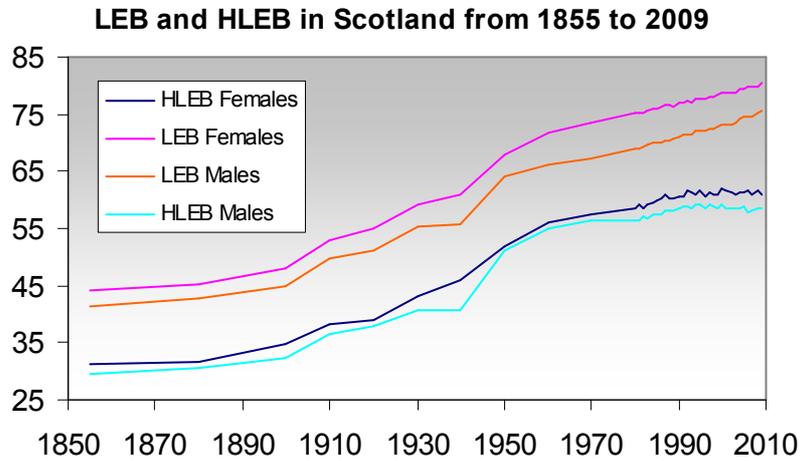

Fig. 8. Estimates for LEB and HLEB for Scotland (1855-2009)

As our direct method for estimating the healthy life expectancy is easy to apply provided that the death and population data are available we have estimated the life expectancy at birth (LEB) and the healthy life expectancy at birth (HLEB) for males and females in Scotland for the time period 1855-2009. The results are included in Tables V and VI . Figure 8 illustrates the LEB and HLEB for total causes for males and females. The results verify the argument for a slower increase of HLEB in nowadays compared to LEB for both males and females. Females have higher LEB and HLEB. However, in the later case the gap is smaller than for LEB. The main interesting point for policy makers is that HLEB is tending to stationarity. This is mainly due to the relative stability of the year for the maximum deterioration age for both males and females for long time periods.

| TABLE VI (Application in Scotland for males) | | | | | | | | | | | |
|---|---|---|---|---|---|---|---|---|---|---|---|
| Males | Max Det | LHLY$_1$ | LHLY$_3$ - LHLY$_1$ | LHLY Light | LHLY Total | HLEB Total | HLEB moderate & severe | HLEB severe | LEB | Scotland official estimates | | |
| Year | Age | Severe | Moderate | | | | | | | HLE (LLI) | HLE New | HLE Old |
| 1855 | 67,1 | 6,7 | 5,4 | 0,0 | 11,9 | 29,5 | 29,2 | 34,6 | 41,4 | | | |
| 1880 | 69,9 | 6,6 | 5,3 | 0,0 | 11,9 | 30,7 | 30,7 | 36,1 | 42,6 | | | |
| 1900 | 70,5 | 6,3 | 6,0 | 0,1 | 12,4 | 32,5 | 32,6 | 38,6 | 44,9 | | | |
| 1910 | 69,2 | 6,1 | 5,0 | 2,1 | 13,2 | 36,7 | 38,8 | 43,8 | 49,9 | | | |
| 1920 | 68,9 | 6,4 | 5,4 | 1,6 | 13,4 | 37,8 | 39,5 | 44,9 | 51,2 | | | |
| 1930 | 66,6 | 6,4 | 5,2 | 3,3 | 14,9 | 40,6 | 43,9 | 49,1 | 55,5 | | | |
| 1940 | 67,2 | 7,3 | 6,5 | 1,1 | 14,9 | 40,8 | 41,9 | 48,4 | 55,7 | | | |
| 1950 | 68,5 | 6,4 | 3,7 | 3,0 | 13,1 | 51,0 | 54,0 | 57,7 | 64,1 | | | |
| 1960 | 64,0* | 5,7 | 2,8 | 2,6 | 11,2 | 55,0 | 57,7 | 60,5 | 66,3 | | | |
| 1970 | 64,0* | 6,0 | 2,9 | 1,7 | 10,6 | 56,5 | 58,2 | 61,1 | 67,1 | | | |
| 1980 | 66,5 | 6,2 | 2,9 | 3,4 | 12,5 | 56,4 | 59,8 | 62,8 | 68,9 | 57,9 | | 62,6 |
| 1981 | 67,6 | 6,3 | 2,9 | 3,4 | 12,6 | 56,5 | 59,9 | 62,8 | 69,1 | 58,4 | | 62,8 |
| 1982 | 66,0 | 6,0 | 2,9 | 3,3 | 12,3 | 57,0 | 60,3 | 63,2 | 69,2 | 57,5 | | 63,7 |
| 1983 | 67,1 | 6,0 | 2,6 | 4,4 | 12,9 | 56,7 | 61,1 | 63,6 | 69,6 | 57,1 | | 63,9 |
| 1984 | 61,0 | 5,5 | 2,5 | 4,6 | 12,6 | 57,3 | 61,9 | 64,4 | 69,9 | 58,3 | | 63,7 |
| 1985 | 61,0 | 5,5 | 2,4 | 4,7 | 12,6 | 57,5 | 62,2 | 64,6 | 70,0 | 58,7 | | 64,3 |
| 1986 | 67,0 | 5,8 | 2,5 | 4,6 | 12,8 | 57,3 | 61,9 | 64,3 | 70,1 | 57,6 | | 64,2 |
| 1987 | 64,5 | 5,6 | 2,5 | 4,2 | 12,3 | 58,1 | 62,4 | 64,9 | 70,5 | 56,9 | | 65,0 |
| 1988 | 67,5 | 5,9 | 2,5 | 3,9 | 12,3 | 58,1 | 62,0 | 64,5 | 70,4 | 56,0 | | 64,6 |
| 1989 | 65,9 | 6,0 | 2,5 | 3,8 | 12,3 | 58,3 | 62,1 | 64,6 | 70,6 | 57,8 | | 65,3 |
| 1990 | 64,1 | 5,8 | 2,5 | 4,3 | 12,6 | 58,5 | 62,8 | 65,3 | 71,1 | 57,3 | | 65,7 |
| 1991 | 65,0 | 6,0 | 2,4 | 4,3 | 12,6 | 58,7 | 63,0 | 65,4 | 71,3 | 59,5 | | 65,6 |
| 1992 | 66,2 | 6,3 | 2,3 | 4,0 | 12,6 | 58,9 | 62,9 | 65,2 | 71,6 | 57,7 | | 66,0 |
| 1993 | 65,6 | 5,8 | 2,8 | 4,3 | 12,9 | 58,5 | 62,7 | 65,5 | 71,4 | 56,0 | | 64,4 |
| 1994 | 67,9 | 5,9 | 2,0 | 5,0 | 12,9 | 59,2 | 64,2 | 66,2 | 72,1 | 58,2 | | 64,6 |
| 1995 | 68,6 | 5,5 | 2,4 | 5,1 | 12,9 | 59,1 | 64,2 | 66,6 | 72,1 | 59,6 | 55,2 | 64,7 |
| 1996 | 70,8 | 5,6 | 2,6 | 5,1 | 13,4 | 58,6 | 63,7 | 66,3 | 72,0 | 57,7 | | 65,7 |
| 1997 | 70,4 | 5,9 | 2,4 | 4,9 | 13,2 | 59,3 | 64,2 | 66,6 | 72,5 | | | |
| 1998 | 69,3 | 5,7 | 2,3 | 5,7 | 13,7 | 58,9 | 64,6 | 66,9 | 72,6 | 60,1 | 57,1 | 65,2 |
| 1999 | 71,5 | 6,1 | 2,5 | 5,7 | 14,3 | 58,4 | 64,1 | 66,5 | 72,7 | | | 64,8 |
| 2000 | 72,3 | 6,4 | 2,3 | 5,2 | 14,0 | 59,1 | 64,4 | 66,7 | 73,1 | 58,9 | | 65,4 |
| 2001 | 72,9 | 6,0 | 2,0 | 6,8 | 14,9 | 58,4 | 65,2 | 67,3 | 73,3 | | | 66,3 |
| 2002 | 73,3 | 6,1 | 2,0 | 6,7 | 14,8 | 58,4 | 65,1 | 67,1 | 73,2 | | | 65,6 |
| 2003 | 73,0 | 6,4 | 2,1 | 6,5 | 15,1 | 58,5 | 65,1 | 67,2 | 73,6 | | 57,3 | |
| 2004 | 73,7 | 6,5 | 1,9 | 7,3 | 15,6 | 58,5 | 65,7 | 67,6 | 74,1 | | | |
| 2005 | 73,5 | 6,9 | 1,9 | 6,9 | 15,6 | 58,9 | 65,8 | 67,7 | 74,6 | | | 67,1 |
| 2006 | 74,5 | 6,4 | 1,9 | 8,6 | 16,9 | 57,8 | 66,4 | 68,2 | 74,7 | | | 67,7 |
| 2007 | 74,3 | 6,5 | 1,8 | 8,3 | 16,6 | 58,1 | 66,4 | 68,2 | 74,7 | | | 67,9 |
| 2008 | 74,8 | 6,6 | 2,0 | 7,9 | 16,6 | 58,5 | 66,5 | 68,5 | 75,1 | | 59,9 | 68,1 |
| 2009 | 75,4 | 6,5 | 1,8 | 8,8 | 17,1 | 58,6 | 67,4 | 69,2 | 75,7 | | 60,0 | |

\* Estimated by a different method

**Conclusions**

What was proven is that both the WHO and EU methods for estimating healthy life expectancy provide important indicators. The WHO system takes into account the severe and moderate causes in estimating the loss of healthy life years; instead the EU system calculates the total loss of healthy life years. For both cases we have proposed a methodology which provides both estimators from only death and population data based on the health state theory of a population that has been extensively described, analysed and applied in this paper and the references included.